\documentclass[twocolumn,preprintnumbers,amsmath,amssymb,superscriptaddress,floatfix]{revtex4-2}

\usepackage{graphicx}
\usepackage{dcolumn}
\usepackage{bm}
\usepackage{color}

\font\scripti=cmmi7
\font\scriptscripti=cmmi5
\def\sib#1{\setbox0 = \hbox{\scripti #1}
  \kern-.02em\copy0\kern-\wd0
  \kern.04em\box0} 
\def\ssib#1{\setbox0 = \hbox{\scriptscripti #1}
  \kern-.02em\copy0\kern-\wd0
  \kern.04em\box0} 
\font\tenib=cmmib10 
\skewchar\tenib='177 \skewchar\tenib='177 \skewchar\tenib='177
\def\pbold#1{\setbox0 = \hbox{$ #1 $}
  \kern-.022em\copy0\kern-\wd0
  \kern.011em\copy0\kern-\wd0
  \kern.011em\copy0\kern-\wd0
  \kern.011em\copy0\kern-\wd0
  \kern.011em\box0} 

\usepackage{graphicx}
\usepackage{dcolumn}
\usepackage{bm}

\def\lesssim{\ \raise.3ex\hbox{$<$}\kern-0.8em\lower.7ex\hbox{$\sim$}\ }
\def\gesim{\ \raise.3ex\hbox{$>$}\kern-0.8em\lower.7ex\hbox{$\sim$}\ }

\newcommand{\red}[1]{{#1}}

\begin{document}
\preprint{RIKEN-iTHEMS-Report-23}
\title{Brueckner $G$-matrix approach to two-dimensional Fermi gases \\ with the finite-range attractive interaction}
\author{Hikaru Sakakibara}
\affiliation{Department of Physics, Graduate School of Science, The University of Tokyo, Tokyo 113-0033, Japan}
\affiliation{Interdisciplinary Theoretical and Mathematical Sciences Program (iTHEMS), RIKEN, Wako, Saitama 351-0198, Japan}
\author{Hiroyuki Tajima}
\affiliation{Department of Physics, Graduate School of Science, The University of Tokyo, Tokyo 113-0033, Japan}
\affiliation{RIKEN Nishina Center, Wako 351-0198, Japan}
\author{Haozhao Liang}
\affiliation{Department of Physics, Graduate School of Science, The University of Tokyo, Tokyo 113-0033, Japan}
\affiliation{Interdisciplinary Theoretical and Mathematical Sciences Program (iTHEMS), RIKEN, Wako, Saitama 351-0198, Japan}

\date{\today}
\begin{abstract}
Two-dimensional spin-1/2 fermions with the finite-range interaction are theoretically studied.
Characterizing the attractive interaction in terms of the scattering length and the effective range,
we discuss the finite-range effects on the ground-state properties in this system.
The Brueckner $G$-matrix approach is employed to analyze the finite-range effects on an attractive Fermi-polaron energy and the equation of state throughout the BEC-BCS crossover in two dimensions, which can be realized in the population-imbalanced and -balanced cases between two components, respectively.
The analytical formulae for these ground states obtained in this study would be useful for understanding many-body phenomena with finite-range interactions in low-dimensional systems.
\end{abstract}
\maketitle
\section{Introduction}
\label{sec:1}

Quantum many-body physics is one of the most exciting and important concepts in various fields of modern physics. 
The inter-particle interaction leads to various quantum phenomena such as superconductivity and superfluidity.
Moreover, the interaction effects become remarkably important in low-dimensional quantum systems.

One of fascinating phenomena associated with strong interactions is the crossover from the molecular Bose-Einstein condensate (BEC) to the Bardeen-Cooper-Schrieffer (BCS) superfluid~\cite{chen2005bcs,zwerger2011bcs,randeria2014crossover,strinati2018bcs,ohashi2020bcs} with changing the attractive interaction by utilizing the Feshbach resonance in cold atoms~\cite{RevModPhys.82.1225}, where tightly-bound dimers gradually change into loosely-bound Cooper pairs without any phase transitions. 
In addition, the BEC-BCS crossover phenomena in two-dimensional (2D) Fermi gases have also been realized by confining the gas in the 2D trap potential~\cite{PhysRevLett.105.030404,PhysRevLett.114.230401,PhysRevLett.120.060402}.
Recently, such strongly-interacting Fermi gases are regarded as useful reference systems to study strong-coupling phenomena in a systematic way~\cite{RevModPhys.80.885,RevModPhys.80.1215}.
Various physical quantities such as equation of state have been measured precisely in the entire BEC-BCS-crossover regime in 2D~\cite{PhysRevLett.106.105301,PhysRevLett.116.045302,holten2022observation,doi:10.1126/science.abc8793,PhysRevLett.128.100401}.

Moreover, the BEC-BCS crossover has been realized in low-dimensional superconductors~\cite{Kasahara2014PNAS,hashimoto2020bose,doi:10.7566/JPSJ.89.102002,nakagawa2021gate,PhysRevX.12.011016} by tuning the carrier density.
As being anticipated in pioneering works~\cite{PhysRev.186.456,nozieres1985bose,PhysRevLett.71.3202}, several electron-hole systems such as graphene also provide another platform to study the BEC-BCS crossover in low-dimensional systems recently~\cite{PhysRevLett.110.146803,park2021tunable,liu2022crossover}.
While the interaction induced by the Feshbach resonance in cold atomic systems can usually be characterized by the zero-range contact-type interaction, other strongly-correlated systems generally involve non-local interactions. 
In semiconductor systems, the finite-range interaction called Rytova-Keldysh potential has been considered~\cite{rytova2018screened,keldysh1979coulomb}.
In the slab phase of neutron stars, the dineutron pairing with the finite-range nucelon force under the quasi-two-dimensional confinement has been discussed~\cite{PhysRevC.79.054305}.
In this regard, the finite-range correction is inevitably important in these density-induced BEC-BCS crossover~\cite{shi2022density,PhysRevA.106.043308}.
In addition, it is reported that the effective range plays a crucial role for the reduced quantum anomaly in 2D~\cite{PhysRevLett.122.070401} observed in recent cold-atomic experiments~\cite{PhysRevLett.121.120401,PhysRevLett.121.120402}. 
The optical control of the effective range and scattering length proposed in Refs.~\cite{PhysRevLett.108.010401,PhysRevA.86.063625} may enable us to study the finite-range effects on cold atomic gases systematically in the future experiments.
Incidentally, quantum Monte Carlo \red{(QMC)} simulations have been performed in the presence of small but nonzero effective ranges~\cite{PhysRevLett.106.110403,PhysRevA.93.023602,PhysRevA.101.033601}.

Another useful setup for examining many-body correlations in cold atomic system is an atomic polaron, which can be realized by preparing an atomic mixture with the population imbalance.
In particular, impurity (minority) atoms immersed in the Fermi sea of majority atoms is referred to as Fermi polarons~\cite{Massignan_2014}. 
The realization of attractive and repulsive Fermi polarons~\cite{PhysRevLett.102.230402,koschorreck2012attractive,PhysRevLett.118.083602} leads to the comprehensive understanding of correlation effects in many-body fermionic systems in a quantitative manner.
\red{In this regard, 2D Fermi polarons with the zero-range attraction have been studied by the diagrammatic QMC simulations~\cite{PhysRevB.89.085119,PhysRevB.90.104510}.} 

The finite-range effects on unitary Fermi polarons in 3D has been investigated by the diffusion Monte Carlo simulation~\cite{PhysRevA.104.043313}.
In 2D, repulsive Fermi polarons with finite-range corrections have been studied in Refs.~\cite{PhysRevA.100.023608,PhysRevA.103.063314,PhysRevA.103.L041302}.
Recently, the properties of two-dimensional Fermi polarons have also attracted much interests in layered electron-hole materials~\cite{sidler2017fermi,muir2022interactions}.
The non-locality, namely, the finite-range correction of the interaction in these systems would be important to understand the similarity and the difference from cold atomic polarons.

In this paper, we discuss the finite-range effects on strongly interacting two-component Fermi gases in 2D.
We characterize the finite-range attractive interaction in terms of scattering length and effective range and employ the Brueckner Hartree-Fock approach with the $G$-matrix~\cite{ring2004nuclear} established in many-body nuclear physics.
The ladder-type diagrams in the particle-particle scattering are summed to give the renormalized self-energy shift on the thermodynamic ground-state quantities such as chemical potential and internal energy~\cite{PhysRevA.63.043606,PhysRevA.85.012701,PhysRevA.95.043625}.
For the contact interactions, the analytical formula of the ground-state energy obtained from the $G$-matrix approach shows a good agreement with \red{the QMC results of the BEC-BCS crossover~\cite{PhysRevLett.106.110403} and the experimental results of attractive Fermi polarons~\cite{koschorreck2012attractive}} in 2D~~\cite{PhysRevA.84.033607,KLAWUNN20162650}.
In this paper, we generalize these approaches to the case with the finite-range interaction and apply them to the BEC-BCS crossover and the attractive Fermi polarons in 2D.
We briefly note that we consider the {\it positive} effective range being possibly relevant to condensed-matter systems~\cite{shi2022density,PhysRevA.106.043308}, in contrast to the previous work for the {\it negative} effective range associated with the narrow Feshbach resonance~\cite{PhysRevA.102.013313}. 

This paper is organized as follows.
In Sec.~\ref{sec:2}, we introduce the model Hamiltonian of spin-$1/2$ fermions with the finite-range attractive interaction.
We show the relation between the interaction parameter in the Hamiltonian and the low-energy constants (i.e., the scattering length and the effective range) by considering the two-body $T$-matrix.
In Sec.~\ref{sec:3}, we present the Brueckner $G$-matrix approach and show the results in the BEC-BCS crossover and in the attractive Fermi polaron in 2D.
Finally, we summarize this paper in Sec.~\ref{sec:4}.
For simplicity, we take $\hbar=k_{\rm B}=1$ and \red{the area $A$ is taken to be unity in the thermodynamic limit}.

\section{Model}
\label{sec:2}

We consider a two-component fermions with finite-range interaction described by the Hamiltonian in the momentum space, i.e.,
\begin{align}
\label{eq:H}
    &\hat{H}
    =\sum_{\bm{k},\sigma}(\varepsilon_{\bm{k}}-\mu_{\sigma})c_{\bm{k},\sigma}^\dag c_{\bm{k},\sigma}\cr
    &\ +\sum_{\bm{k},\bm{k}',\bm{P}}
    U(\bm{k},\bm{k}')
    c_{\bm{k}+\frac{\bm{P}}{2},\uparrow}^\dag
    c_{-\bm{k}+\frac{\bm{P}}{2},\downarrow}^\dag
    c_{-\bm{k}'+\frac{\bm{P}}{2},\downarrow}
    c_{\bm{k}'+\frac{\bm{P}}{2},\uparrow},
\end{align}
where $\varepsilon_{\bm{k}}=\frac{k^2}{2m}$ is the kinetic energy of a fermion with a momentum $\bm{k}$ and a mass $m$.
$\mu_{\sigma}$ is the chemical potential for the state with spin $\sigma=\uparrow,\downarrow$.
$c_{\bm{k},\sigma}$ and $c_{\bm{k},\sigma}^\dag$ are the fermionic annihilation and creation operators, respectively.

The second term in Eq.~(\ref{eq:H}) denotes the interaction.
We introduce the separable $s$-wave interaction
\begin{align}
\label{eq:u0}
    U(\bm{k},\bm{k}')=U_0\gamma_{k}\gamma_{k'},
\end{align}
where $U_0$ and $\gamma_k$ are the momentum-independent coupling constant and the form factor, respectively.
Since we are interested in the attractive interaction,
we take $U_0<0$.
In this study, we employ
\begin{align}
\label{eq:gamma}
    \gamma_k=\frac{1}{\sqrt{1+(k/\Lambda)^2}},
\end{align}
which reproduces the relative momentum dependence of the scattering phase shift $\delta_k$ up to $O(k^2)$~\cite{PhysRevA.97.013601,JPSJ.88.093001,PhysRevA.106.043308}.
$\Lambda$ is the cutoff scale and it may be associated with the inverse of screening length in semiconductor systems~\cite{RevModPhys.90.021001}.
A similar form factor $\gamma_k=1/[1+(k/\Lambda)^2]^j$ ($j$ is an integer) 
has been employed in the study of semiconductor systems~\cite{mohseni2022trion}.

To see the relation between the low-energy constants (i.e., the scattering length $a$ and the effective range $R$) and the model parameters (i.e., the coupling constant \red{$U_0$} and the cutoff $\Lambda$), we examine the two-body $T$-matrix given by
\begin{align}
\label{eq:tmat}
T(\bm{k},\bm{k}';\omega)&=U(\bm{k},\bm{k}')+\sum_{\bm{p}}
    \frac{U(\bm{k},\bm{p})T(\bm{p},\bm{k}';\omega)}{\omega_+-2\varepsilon_{\bm{p}}},
\end{align}
where $\omega_{+}=\omega+i\delta$ is the two-body energy with infinitesimally small imaginary part $i\delta$.
Once the separable interaction in Eq.~(\ref{eq:u0}) is considered,
the separable form of the $T$-matrix is obtained as
\begin{align}
\label{eq:tmat2}
T(\bm{k},\bm{k}';\omega)=\gamma_kt(\omega)\gamma_{k'}.
\end{align}
Accordingly, we obtain
\begin{align}
\label{eq:t}
    t(\omega)&=U_0\left[1-U_0\sum_{\bm{p}}
    \frac{\gamma_p^2}{\omega_+-p^2/m}\right]^{-1}\cr
    &\equiv
    U_0\left[1-U_0\Pi(\omega)\right]^{-1}.
\end{align}
The momentum integration in the in-vacuum pair propagator $\Pi(\omega)$ can be evaluated analytically as
\begin{align}
\label{eq:Pi}
    \Pi(\omega)
    &=-\frac{m\Lambda^2}{4\pi(m\omega+\Lambda^2)}\ln\left(-\frac{\Lambda^2}{m\omega_+}\right).
\end{align}
The scattering length and the effective range can be expressed in terms of the on-shell $T$-matrix as~\red{\cite{adhikari1986quantum}}
\begin{align}
T(\bm{k},\bm{k};2\varepsilon_{\bm{k}})
=\frac{4\pi}{m}\left[-2\ln(ka)-R^2k^2+i\pi\right]^{-1}.
\end{align}
Using Eqs.~\eqref{eq:t} and \eqref{eq:Pi},
we obtain
\begin{align}
a=\frac{1}{\Lambda}\exp\left(-\frac{2\pi}{mU_0}\right),
\end{align}
\begin{align}
\label{eq:R}
    R^2=-\frac{4\pi}{mU_0\Lambda^2}>0.
\end{align}
In this regard, we find the dimensionless quantity
\begin{align}
    \frac{R^2}{a^2}=-\frac{4\pi}{mU_0}
    \exp\left(\frac{4\pi}{mU_0}\right).
\end{align}
In the present case, there are no higher-order coefficients $O(k^4)$ such as shape parameter~\cite{PhysRevA.106.043308}.
In this paper,
we consider the positive $R^2$ while the negative $R^2$ has been investigated in Ref.~\cite{PhysRevA.96.023619}.
To generalize the present result with both positive and negative $R^2$ cases,
it is necessary to use the two-channel model with the form factor~\cite{JPSJ.88.093001}, which is beyond the present scope.

Moreover, it is known that the two-body bound state is present even for arbitrary attractive short-range interaction strength in 2D.
The two-body binding energy $E_{\rm b}$ is obtained from a pole of the $T$-matrix as
\begin{align}
    1-U_0\Pi(\omega=-E_{\rm b})=0.
\end{align}
More explicitly, one gets
\begin{align}
    \frac{4\pi}{mU_0}&=\frac{1}{1-\frac{mE_{\rm b}}{\Lambda^2}}\ln\left(\frac{mE_{\rm b}}{\Lambda^2}\right).
\end{align}
We note that $E_{\rm b,0}=1/(ma^2)$ is obtained in the zero-range limit ($R\rightarrow 0$, corresponding to $\Lambda\rightarrow\infty$)~\cite{levinsen2015strongly}.
\begin{figure}[t]
    \centering
    \includegraphics[width=8cm]{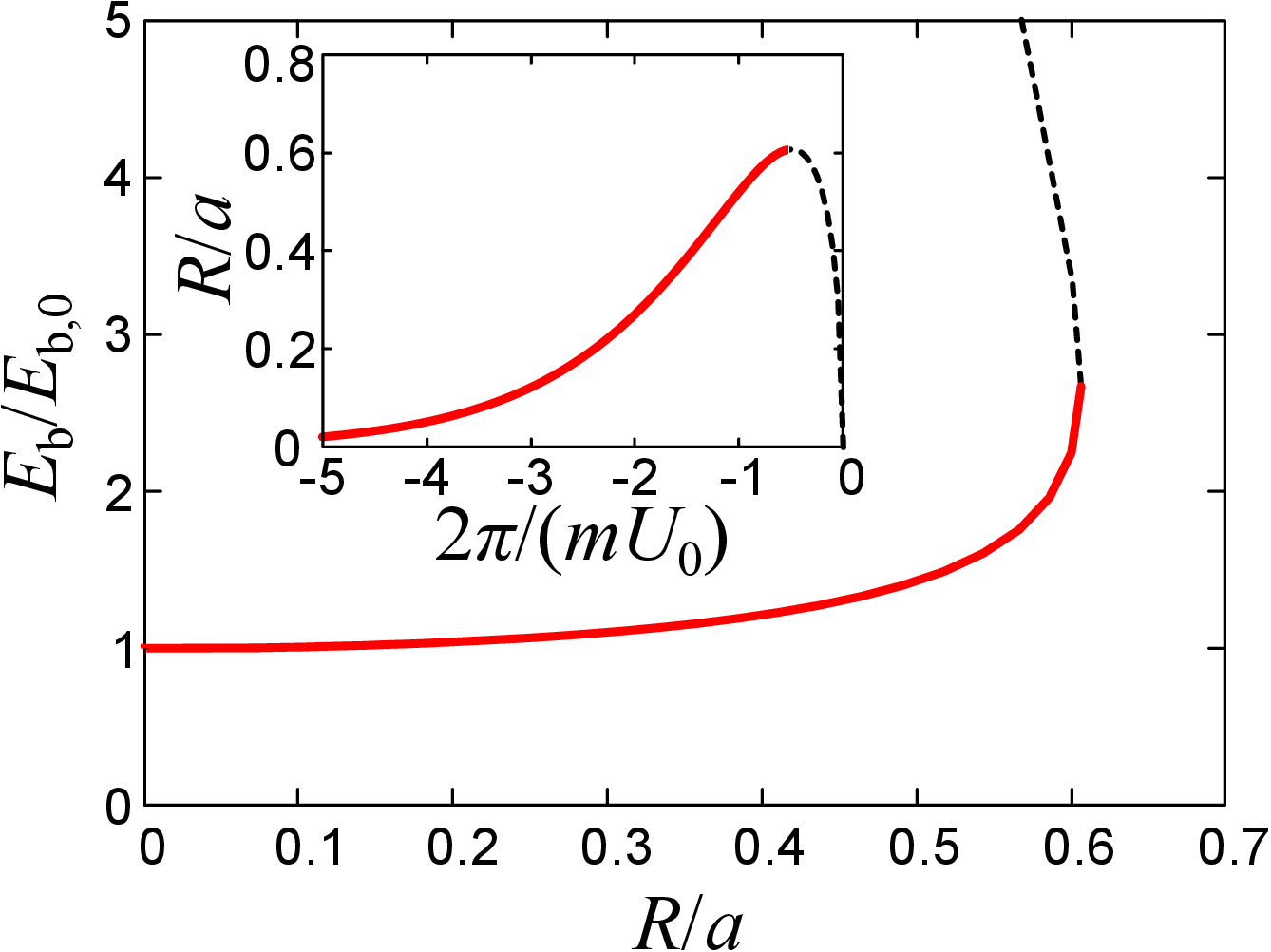}
    \caption{Two-body binding energy $E_{\rm b}$ as a function of the ratio between the effective range $R$ and the scattering length $a$. $E_{\rm b,0}=1/(ma^2)$ is the binding energy at $R=0$. The inset shows $R/a$ as a function of the inverse coupling constant $\frac{2\pi}{mU_0}$. The red solid line corresponds to the parameter regime $E_{\rm b}\leq \frac{\Lambda^2}{m}$, where we consider in this paper. We also plot the region where a deep bound state is found ($E_{\rm b}>\frac{\Lambda^2}{m}$) with the black dashed line.}
    \label{fig:1}
\end{figure}

To examine the finite-range correction on the binding energy, in Fig~\ref{fig:1} we plot
\begin{align}
    \frac{E_{\rm b}}{E_{\rm b,0}}&=
    \frac{mE_{\rm b}}{\Lambda^2}\exp\left(-\frac{4\pi}{mU_0}\right).
\end{align}
If one increases $R$,
$E_{\rm b}$ becomes larger compared to the zero-range result $E_{\rm b,0}$ for a given scattering length $a$.
On the one hand, a similar dimensionless quantity has been shown in Ref.~\cite{PhysRevA.101.033601} and the behavior of the binding energy with increasing the effective range is consistent with the present result.
On the other hand, two solutions of $E_{\rm b}/E_{\rm b,0}$ can be found at the strong coupling side (i.e., $\frac{2\pi}{mU_0}\geq -\frac{1}{2}$ and $E_{\rm b}\geq \frac{\Lambda^2}{m}$).
Such a situation can also be found in the case with a positive scattering length and nonzero effective range~\cite{naidon2017efimov}, which is also known as a spurious pole~\cite{ebert2021alternative}.
In this paper, we focus on the weak-coupling parameter region, where $\frac{2\pi}{mU_0}<-\frac{1}{2}$ and $R/a<e^{-1/2}\simeq 0.607$ (i.e., the red solid line in Fig.~\ref{fig:1}).

\section{Brueckner G-matrix approach}
\label{sec:3}
In this section, we derive the effective interaction based on the Brueckner $G$-matrix approach developed in many-body nuclear physics~\cite{ring2004nuclear}.
For the in-medium two-body scattering, we introduce the Pauli-blocking projection $Q(\bm{p},\bm{P})$ in the momentum integral of the $G$-matrix, 
\begin{align}
    &\red{G(\bm{k},\bm{k}';\bm{P},\omega)}=U(\bm{k},\bm{k}')\cr
    &\quad\quad+\sum_{\bm{p}}
    \frac{U(\bm{k},\bm{p})Q(\bm{p},\bm{P})\red{G(\bm{p},\bm{k}';\bm{P},\omega)}}{\omega_+-p^2/m-P^2/(4m)},
\end{align}
where the intermediate state is restricted by the Pauli-blocking effect.
Similar to the two-body $T$-matrix with the separable interaction in Eq.~(\ref{eq:tmat2}), we obtain the separable form of the $G$-matrix $G(\bm{k},\bm{k}';\bm{P},\omega)=\gamma_{k}g(\bm{P},\omega)\gamma_{k'}$, where 
\begin{align}
\label{eq:g}
    g(\bm{P},\omega)
    =\left[\frac{1}{U_0}-\sum_{\bm{p}}\frac{\gamma_p^2Q(\bm{p},\bm{P})}{\omega_+-p^2/m-P^2/(4m)}\right]^{-1}.
\end{align}

The explicit form of $Q(\bm{p},\bm{P})$ within the
Tamm-Dancoff approximation (TDA)~\cite{ring2004nuclear} is given by
\begin{align}
\label{eq:Q}
    Q(\bm{p},\bm{P})=\theta(|\bm{P}/2+\bm{p}|-k_{\rm F})\theta(|\bm{P}/2-\bm{p}|-k_{\rm F}),
\end{align}
where fermions below $E_{\rm F}$ are suppressed due to the Pauli-blocking effect.
\red{In more details, Eq.~\eqref{eq:Q} represents the two-particle distributions above the Fermi sea in the intermediate state as $Q(\bm{p},\bm{P})=[1-f(\varepsilon_{\bm{p}+\bm{P}/2}-\mu_{\rm \uparrow})][1-f(\varepsilon_{-\bm{p}+\bm{P}/2}-\mu_{\rm \downarrow})]$,
where $f(x)=\theta(-x)$ is the Fermi distribution function at $T=0$.
However, the two-hole distributions below the Fermi sea $-f(\varepsilon_{\bm{p}+\bm{P}/2}-\mu_{\rm \uparrow})f(\varepsilon_{-\bm{p}+\bm{P}/2}-\mu_{\rm \downarrow})$ is neglected, while the many-body $T$-matrix approach~\cite{ohashi2020bcs} includes both effects as
$Q(\bm{p},\bm{P})=[1-f(\varepsilon_{\bm{p}+\bm{P}/2}-\mu_{\rm \uparrow})][1-f(\varepsilon_{-\bm{p}+\bm{P}/2}-\mu_{\rm \downarrow})]-f(\varepsilon_{\bm{p}+\bm{P}/2}-\mu_{\rm \uparrow})f(\varepsilon_{-\bm{p}+\bm{P}/2}-\mu_{\rm \downarrow})\equiv 1-f(\varepsilon_{\bm{p}+\bm{P}/2}-\mu_{\rm \uparrow})-f(\varepsilon_{-\bm{p}+\bm{P}/2}-\mu_{\rm \downarrow})$.
In the context of the variational calculations, TDA corresponds to the Cooper problem and is qualitatively valid in the BEC-BCS crossover regime~\cite{PhysRevA.86.013628,PhysRevA.104.053328}.
Moreover, 
Compared to the BEC-BCS crossover in population-balanced Fermi gases,
TDA gives better description of many-body effects in the Fermi polaron problem where the minority distribution $f(\varepsilon_{-\bm{p}+\bm{P}/2}-\mu_{\rm \downarrow})$ is taken to be zero and hence the two-hole distribution vanishes.
}

\subsection{Equation of state in 2D BEC-BCS crossover}

First, we examine the equation of state in the 2D BEC-BCS crossover in population-balanced Fermi gases, where $\mu_{\uparrow}=\mu_{\downarrow}\equiv \mu$ and $N_{\uparrow}=N_{\downarrow}$
($N_{\sigma}$ is the number of the state $\sigma$).
The ground-state energy $E$ is given by the expectation value of the canonical Hamiltonian, 
\begin{align}
    E&=\langle \hat{H}\rangle +\mu N\equiv E_{\rm FG}+\langle \hat{V}\rangle,
\end{align}
where we decomposed $E$ into the free-gas part $E_{\rm FG}=\frac{1}{2}NE_{\rm F}$ and the interaction energy $\langle \hat{V}\rangle$ by following the procedure in Ref.~\cite{KLAWUNN20162650}. 
Here, $N=\sum_{\sigma}N_{\sigma}=\frac{k_{\rm F}^2}{2\pi}$ and $E_{\rm F}=\frac{k_{\rm F}^2}{2m}$ are the total fermion number and the Fermi energy, respectively, where $k_{\rm F}$ is the corresponding Fermi momentum.
Assuming that the momentum- and energy-dependent effective interaction in medium is dominated by the component at the zero center-of-mass momentum ($\bm{P}\simeq\bm{0}$) and at the bound-state energy ($\omega=-E_{\rm b}$)~\cite{KLAWUNN20162650},
we evaluate $\langle \hat{V}\rangle$ approximately by using the $G$-matrix as
\begin{align}
\label{eq:V}
    \langle \hat{V}\rangle
    &\equiv
    \sum_{\bm{k},\bm{k}',\bm{P}}
    U(\bm{k},\bm{k}')\cr
    &\times\langle c_{\bm{k}+\bm{P}/2,\uparrow}^\dag
    c_{-\bm{k}+\bm{P}/2,\downarrow}^\dag
    c_{-\bm{k}'+\bm{P}/2,\downarrow}
    c_{\bm{k}'+\bm{P}/2,\uparrow}\rangle\cr
    &\simeq
    \sum_{\bm{k},\bm{k}'}G(k,k',\bm{P}\simeq\bm{0},\omega\simeq-E_{\rm b})
    \langle c_{\bm{k},\uparrow}^\dag
    c_{\bm{k},\uparrow}\rangle
    \langle
    c_{\bm{k}',\downarrow}^\dag
    c_{\bm{k}',\downarrow}
    \rangle\cr
    &=\sum_{\bm{k},\bm{k}'}g(\bm{0},-E_{\rm b})\gamma_{k}\gamma_{k'}\theta(k_{\rm F}-k)\theta(k_{\rm F}-k'),
\end{align}
where we took the Fermi-Dirac distribution $\langle c_{\bm{k},\sigma}^\dag c_{\bm{k},\sigma}\rangle\simeq \theta(k_{\rm F}-k)$.
\red{In the second line of Eq.~\eqref{eq:V},
the four-point correlation is evaluated as
$\langle c_{\bm{k}+\bm{P}/2,\uparrow}^\dag
    c_{-\bm{k}+\bm{P}/2,\downarrow}^\dag
    c_{-\bm{k}'+\bm{P}/2,\downarrow}
    c_{\bm{k}'+\bm{P}/2,\uparrow}\rangle\simeq 
        \langle c_{\bm{k},\uparrow}^\dag
    c_{\bm{k},\uparrow}\rangle
    \langle
    c_{\bm{k}',\downarrow}^\dag
    c_{\bm{k}',\downarrow}
    \rangle
    $, which enables us to evaluate the interaction energy in an analytical way.
Moreover, the bare coupling constant $U(\bm{k},\bm{k}')$ is replaced by the $G$-matrix $G(k,k',\bm{P}\simeq\bm{0},\omega\simeq-E_{\rm b})$ in the low-energy limit to include strong-coupling effects associated with the bound-state formation.}
Although
this approximation may lack several important microscopic properties, such as the effects of the pairing gap and the dimer-dimer interaction,
it is still useful to understand the macroscopic ground-state properties of the system in an analytical way.
Indeed, the analytical form of the equation of state based on this approach~\cite{KLAWUNN20162650} shows a good agreement with the \red{QMC} results in 2D Fermi gases \red{with the small effective range}~\cite{PhysRevLett.106.110403}.
The momentum summation in Eq.~\eqref{eq:V} can be performed analytically as
\begin{align}
    \langle \hat{V}\rangle
    &\simeq \int_0^{k_{\rm F}}\frac{d^2\bm{k}}{(2\pi)^2}
    \int_0^{k_{\rm F}}\frac{d^2\bm{k}'}{(2\pi)^2}
    g(\bm{0},-E_{\rm b})\gamma_{k}\gamma_{k'}
    \cr
    &=\frac{g(\bm{0},-E_{\rm b})\Lambda^2}{4\pi^2}
    \left(\sqrt{\Lambda^2+k_{\rm F}^2}-\Lambda\right)^2.
\end{align}
Moreover, taking $\bm{P}=\bm{0}$ and $\omega=-E_{\rm b}$ in Eq.~\eqref{eq:g},
we obtain
\begin{align}
\label{eq:g2}
    g(\bm{0},-E_{\rm b})&=\left[\frac{1}{U_0}-\sum_{\bm{p}}\frac{\theta(p-k_{\rm F})\gamma_p^2}{-E_{\rm b}-p^2/m}\right]^{-1}\cr
    &=\frac{4\pi}{m}
    \frac{1-\frac{mE_{\rm b}}{\Lambda^2}}{
    \ln\left(1+\frac{k_{\rm F}^2}{\Lambda^2}\right)
    -
    \ln\left(1+\frac{k_{\rm F}^2}{mE_{\rm b}}\right)}.
\end{align}
In this way, the interaction energy is given by
\begin{align}
    \langle \hat{V}\rangle=\frac{\Lambda^2}{\pi m}
    \frac{\left(\sqrt{\Lambda^2+k_{\rm F}^2}-\Lambda\right)^2\left(1-\frac{mE_{\rm b}}{\Lambda^2}\right)}{
    \ln\left(1+\frac{k_{\rm F}^2}{\Lambda^2}\right)
    -
    \ln\left(1+\frac{k_{\rm F}^2}{mE_{\rm b}}\right)}.
\end{align}
At $\Lambda\rightarrow \infty$, we find
\begin{align}
    \langle \hat{V}\rangle&\simeq
    -E_{\rm F}N\frac{1}{\ln\left(1+\frac{2E_{\rm F}}{E_{\rm b,0}}\right)},
\end{align}
which reproduces the zero-range result in Ref.~\cite{KLAWUNN20162650}.
Moreover, in the zero-range weak-coupling limit ($E_{\rm F}\gg E_{\rm b,0}$, corresponding to $\frac{2E_{\rm F}}{E_{\rm b,0}}\equiv k_{\rm F}^2a^2\gg 1$), we recover the weak-coupling formula $E \simeq \frac{1}{2}E_{\rm F}N\left[1-\frac{1}{\ln(k_{\rm F}a)}\right]$~\cite{PhysRevB.12.125,PhysRevA.84.063626}.
In the deep BEC limit where $E_{\rm b,0}\gg E_{\rm F}$, one can also obtain $E\simeq -\frac{N E_{\rm b,0}}{2}$, corresponding to the binding energy of $N/2$ molecules as expected. 
Accordingly, the dimensionless form of the ground-state energy $E/E_{\rm FG}$ is given by
\begin{align}
    \frac{E}{E_{\rm FG}}
    &=1+\frac{8\Lambda^2}{ k_{\rm F}^2}
    \frac{\left(\sqrt{1+\frac{\Lambda^2}{k_{\rm F}^2}}-\frac{\Lambda}{k_{\rm F}}\right)^2
    \left(1-\frac{E_{\rm b}}{2E_{\rm F}}\frac{k_{\rm F}^2}{\Lambda^2}\right)}{\ln\left(1+\frac{k_{\rm F}^2}{\Lambda^2}\right)-\ln\left(1+\frac{2E_{\rm F}}{E_{\rm b}}\right)}.
\end{align}

\begin{figure}[t]
    \centering
    \includegraphics[width=8cm]{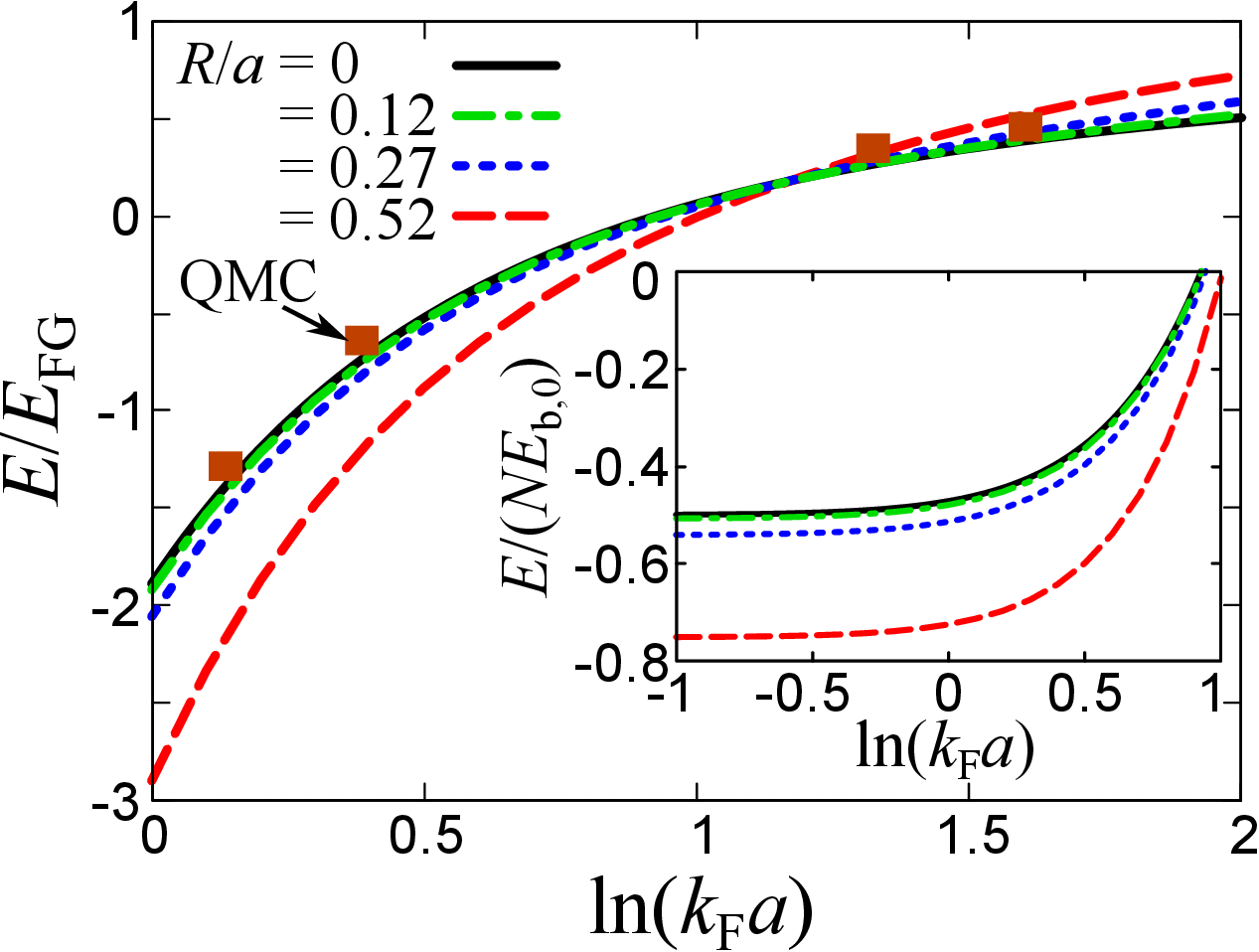}
    \caption{Internal energy $E/E_{\rm FG}$ in the 2D BEC-BCS crossover with different effective ranges, $R/a=0$, $0.12$, $0.27$, and $0.52$. $E_{\rm FG}=\frac{1}{2}NE_{\rm F}$ is the ground-state energy in an ideal Fermi gas. The inset shows the energy per particle $E/N$ normalized by $E_{\rm b,0}$, where $E_{\rm b,0}=1/(ma^2)$ is the two-body binding energy at the zero-range limit. 
    \red{For comparison, the QMC results with $k_{\rm F}R=0.0025$~\cite{PhysRevLett.106.110403} are shown (square). Note that the definition of $a$ is different from Ref.~\cite{PhysRevLett.106.110403}.}}
    \label{fig:2}
\end{figure}
Figure~\ref{fig:2} shows $E/E_{\rm FG}$ in the BEC-BCS crossover with different $R/a$. 
\red{First, one can confirm that the zero-range result ($R/a=0$) well reproduces the QMC result~\cite{PhysRevLett.106.110403}.
While this QMC result involves the finite-range correction with $k_{\rm F}R=0.0025$, this value gives a sufficiently small ratio $0.00033<R/a<0.0025$ in the range of $0\leq \ln(k_{\rm F}a)\leq 2$.
}
In the high-density weak-coupling regime [i.e., $\ln(k_{\rm F}a)\gesim 1$], the effective-range correction (in other words, the finite-cutoff correction) reduces the magnitude of the interaction energy $\langle \hat{V}\rangle$ as
\begin{align}
    \langle \hat{V}\rangle
    \simeq\frac{g(\bm{0},-E_{\rm b})k_{\rm F}^4}{16\pi^2}
    \left(1-\frac{k_{\rm F}^2}{2\Lambda^2}\right)+O(k_{\rm F}^4/\Lambda^4),
\end{align}
regardless of the enhanced effective coupling $g(\bm{0},-E_{\rm b})\simeq -\frac{4\pi}{m}\frac{1}{\ln(k_{\rm F}a)-\frac{k_{\rm F}^2}{\Lambda^2}}$ in the weak-coupling limit. 
We note that the expansion with respect to $k_{\rm F}/\Lambda$ is equivalent to that of $k_{\rm F}R\sqrt{\frac{m|U_0|}{4\pi}}$ based on Eq.~\eqref{eq:R}.
Since $k_{\rm F}R\rightarrow 0$ is realized in the low-density limit, the correction proportional $k_{\rm F}/\Lambda$ is negligible in this regime.
However, in the low-density strong-coupling regime [i.e., $\ln(k_{\rm F}a)\lesssim 1$],
$E$ becomes negative and is strongly reduced by the effective-range correction.
This is due to the enlargement of $E_{\rm b}$ as shown in Fig.~\ref{fig:1}.
To see this, it is useful to see the energy per particle $E/N$ scaled by a fixed energy scale (where $E_{\rm b,0}$ is adopted in this paper) as
\begin{align}
    \frac{E}{NE_{\rm b,0}}\equiv\frac{E}{E_{\rm FG}}\frac{E_{\rm FG}}{NE_{\rm b,0}}
    \equiv\frac{E}{E_{\rm FG}}\frac{k_{\rm F}^2a^2}{4},
\end{align}
which is shown in the inset of Fig.~\ref{fig:2}.
The flat region of $E/NE_{\rm b,0}$ in the density dependence can be found at $\ln(k_{\rm F}a)\lesssim 0$,
indicating that the energy per particle is describe as $E/N=-\frac{1}{2}E_{\rm b}$.
Indeed, the zero-range result approaches $E/NE_{\rm b,0}=-1/2$ as expected.
In the presence of nonzero $R$, $E/NE_{\rm b,0}$ approaches
$-\frac{1}{2}\frac{E_{\rm b}}{E_{\rm b,0}}\leq -\frac{1}{2}$.
In this way, one can see the reduction of $E$ due to the larger $E_{\rm b}$ with increasing $R/a$ in the low-density regime.
We also note that $E/NE_{\rm b,0}$ shown in the inset of Fig.~\ref{fig:2} is useful to see the stability of the system towards the density collapse.
Even in the presence of the nonzero effective range, the system is found to be stable within the present approach in contrast to the previous work in 3D~\cite{PhysRevA.95.013633},
\red{which studied the larger effective-range regime than that of Ref.~\cite{PhysRevA.86.053603}.}

\subsection{2D attractive Fermi polaron}
In this subsection, we calculate the attractive Fermi polaron energy within the Bruckner $G$-matrix approach.
The Fermi polaron system can be realized in the two-component mixture with the large population imbalance $\frac{N_{\downarrow}}{N_{\uparrow}}\ll 1$.
While many Fermi polarons exist (i.e., $1\ll N_{\downarrow}\ll N_{\uparrow}$) in actual experiments,
we take $N_{\downarrow}=1$ for convenience.
This assumption works relatively well in the Fermi polaron problem~\cite{tajima2018many,PhysRevA.98.013626,tajima2021polaron}.
Moreover, one can examine the validity of the present approach on the self-energy shift without concerning the existence of the pairing gap.
As shown in Ref.~\cite{PhysRevA.84.033607},
the attractive Fermi polaron energy $E_{\rm P}$ is given by the Brueckner Hartree-Fock self-energy at zero momentum as
\begin{align}
\label{eq:epolaron}
    E_{\rm P}&= \Sigma_{\downarrow}(\bm{p}=\bm{0})\cr
    &\equiv\sum_{\bm{k}}G(\bm{k},\bm{k};\bm{0},-E_{\rm b})\theta(k_{\rm F,\uparrow}-k),
\end{align}
where $k_{\rm F,\uparrow}=\sqrt{4\pi N_{\uparrow}}$ is the Fermi momentum of the majority component.
We note that in this subsection $\mu_{\uparrow}=E_{\rm F,\uparrow}$
and $\mu_{\downarrow}=E_{\rm P}$ are taken,
where $E_{\rm F,\uparrow}=\frac{k_{\rm F,\uparrow}^2}{2m}$ is the Fermi energy of the majority component.
Performing the momentum summation in Eq.~\eqref{eq:epolaron} \red{with $Q(\bm{p},\bm{0})=\theta(p-k_{\rm F})$ in Eq.~\eqref{eq:g2}},
one can obtain an analytical expression of $E_{\rm P}$ as
\begin{align}
    E_{\rm P}&=
    \frac{\left(\frac{\Lambda^2}{m}-E_{\rm b}\right)
    \ln\left(1+\frac{k_{\rm F,\uparrow}^2}{\Lambda^2}\right)
    }{\ln\left(1+\frac{k_{\rm F,\uparrow}^2}{\Lambda^2}\right)
    -
    \ln\left(1+\frac{k_{\rm F,\uparrow}^2}{mE_{\rm b}}\right)
    }.
\end{align}
Indeed, in the limit of $\Lambda\rightarrow\infty$, we obtain
\begin{align}
\label{eq:ep}
    E_{\rm P}\simeq  -\frac{2E_{\rm F,\uparrow}}{\ln\left(1+\frac{2E_{\rm F,\uparrow}}{E_{\rm b,0}}\right)}.
\end{align}
Eq.~(\ref{eq:ep}) reproduces the zero-range result in Ref.~\cite{PhysRevA.84.033607}.
The zero-range weak-coupling limit $E_{\rm P}\simeq -2E_{\rm F,\uparrow}/\ln(2E_{\rm F,\uparrow}/E_{\rm b})$ also agrees with that obtained by the variational method~\cite{PhysRevA.83.021603}, while $E_{\rm P}\simeq -E_{\rm b}+E_{\rm F,\uparrow}$ is found in the zero-range strong-coupling limit.
The dimensionless form with the nonzero effective range reads
\begin{align}
    \frac{E_{\rm P}}{E_{\rm F,\uparrow}}
    &=
    \frac{\left(\frac{2\Lambda^2}{k_{\rm F,\uparrow}^2}
    -\frac{E_{\rm b}}{E_{\rm F,\uparrow}}\right)
    \ln\left(1+\frac{k_{\rm F,\uparrow}^2}{\Lambda^2}\right)}{\ln\left(1+\frac{k_{\rm F,\uparrow}^2}{\Lambda^2}\right)-\ln\left(1+\frac{2E_{\rm F,\uparrow}}{E_{\rm b}}\right)}.
\end{align}

\begin{figure}[t]
    \centering
    \includegraphics[width=8cm]{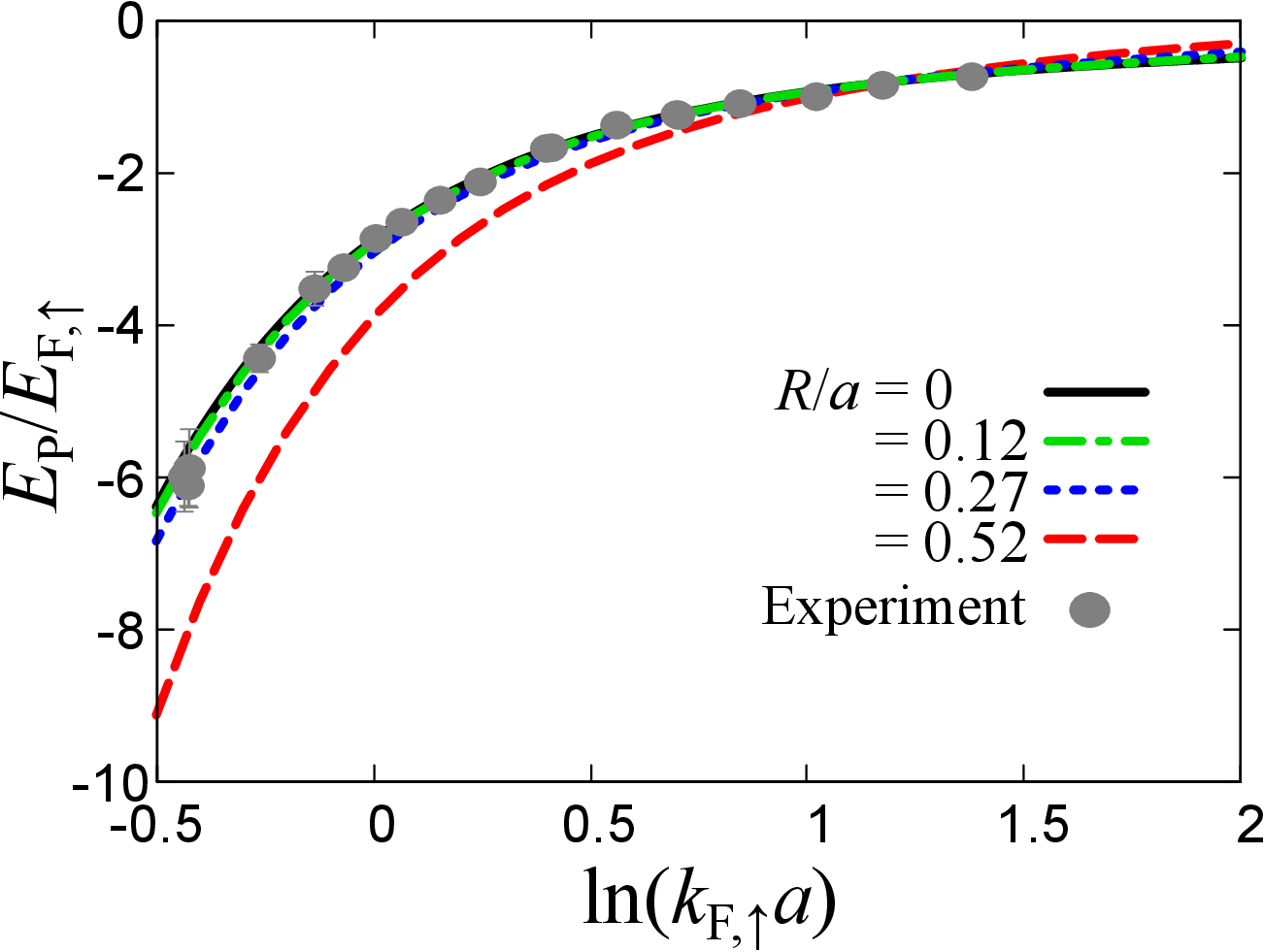}
    \caption{Attractive Fermi polaron energy as a function of $\ln(k_{\rm F}a)$ at different effective ranges, $R/a=0$, $0.12$, $0.27$, and $0.52$.
    \red{The grey dots show the experimental results~\cite{koschorreck2012attractive}, which can be regarded as the zero-range case.}
    }
    \label{fig:3}
\end{figure}

Figure~\ref{fig:3} shows the calculated $E_{\rm P}/E_{\rm F,\uparrow}$ as a function of $\ln(k_{\rm F}a)$ with different $R/a$.
The smooth crossover from the weakly-attractive polaron regime to the strong-coupling one can be found for each $R/a$.
Similar to the result of the internal energy in the 2D BEC-BCS crossover, 
the substantial reduction of $E_{\rm P}$ associated with nonzero $R$ can be found in the low-density strong-coupling regime [i.e., $\ln(k_{\rm F,\uparrow}a)\lesssim 1$], where $E_{\rm b}$ is enlarged by the effective-range correction.
Also, $E_{\rm P}$ slightly increases in the high-density weak-coupling regime [i.e., $\ln(k_{\rm F,\uparrow}a)\gesim 1$] as $E/E_{\rm FG}$ increases
at $\ln(k_{\rm F,\uparrow}a)$ with increasing $R/a$ in Fig.~\ref{fig:2}.
Again such a tendency can be understood from Eq.~\eqref{eq:epolaron}
where $E_{\rm P}=\frac{g(\bm{0},-E_{\rm b})\Lambda^2}{4\pi}\ln\left(1+\frac{k_{\rm F,\uparrow}^2}{\Lambda^2}\right)=-\frac{|g(\bm{0},-E_{\rm b})|k_{\rm F,\uparrow}^2}{4\pi}\left(1-\frac{k_{\rm F,\uparrow}^2}{2\Lambda^2}\right)+O(k_{\rm F,\uparrow}^4/\Lambda^4)$
is found after the momentum summation.
In this regard, the finite cutoff suppresses the interaction near the Fermi surface at high density.

\red{For comparison, the experimental results of the attractive polaron energy~\cite{koschorreck2012attractive} are shown in Fig.~\ref{fig:3}, which can be regarded as the zero-range results.
While the importance of the quasi-2D nature has been pointed out in Ref.~\cite{PhysRevA.87.033616}, the good agreement between the experiment and our zero-range result indicates that our $G$-matrix approach is sufficiently useful to discuss qualitative features of 2D attractive Fermi polarons.}

\section{Summary}
\label{sec:4}

In this paper, effects of the finite-range attractive interaction have been investigated in two-dimensional spin-$1/2$ Fermi gases.
We have employed the separable finite-range interaction, which reproduces the 2D $s$-wave scattering phase shift within the effective-range expansion.
It has been applied to study the finite-range effects on the BEC-BCS crossover and the Fermi polarons in 2D.

Using the Brueckner $G$-matrix approach involving the particle-particle scattering process with the Pauli-blocking effect, we have derived the analytical formula of the equation of state in the BEC-BCS crossover and the attractive Fermi polaron energy in the presence of the nonzero effective range.
It is found that, while the substantial reduction of the internal energy is found in the low-density BEC regime due to the enhanced two-body binding energy with finite-range corrections, the finite-cutoff associated with the effective range suppresses the pairing energy gain in the high-density BCS regime.
A similar effective-range dependence can be found in the attractive Fermi polaron energy with increasing the dimensionless coupling parameter [i.e., $\ln(k_{\rm F,\uparrow}a)$].

For future work,
it would be important to systematically examine the pairing properties with finite-range corrections by using the Brueckner Hartree-Fock-Bogoliubov theory, where both the pairing and density mean-fields are taken into account. 
The present approach can be extended to other systems such as mass-imbalanced mixtures, \red{repulsive gases on the upper branch of the Feshbach resonance}, and electron-hole systems.
Furthermore, the study on the finite-temperature properties and the Berezinskii-Kosterlitz-Thouless transition~\cite{berezinskii1972destruction,Kosterlitz_1973,Kosterlitz_1974} would be an important future direction. 

\acknowledgements
H.S. was supported by RIKEN Junior Research Associate Program.
H.T. acknowledges the JSPS Grants-in-Aid for Scientific Research under Grant Nos.~18H05406, ~22K13981,~22H01158.
H.L. acknowledges the JSPS Grant-in-Aid for Early-Career Scientists under Grant No.~18K13549, the JSPS Grant-in-Aid for Scientific Research (S) under Grant No.~20H05648, and the RIKEN Pioneering Project: Evolution of Matter in the Universe.

\bibliographystyle{apsrev4-2}
\bibliography{reference.bib}

\end{document}